\documentclass[apl,preprint,onecolumn]{revtex4}

\usepackage{graphicx}
\usepackage{dcolumn}
\usepackage{bm}

\begin{document}

\title{Prediction of polar ordered oxynitride perovskites}

\author{Razvan Caracas}
\affiliation{Geophysical Laboratory, Carnegie Institution of
Washington, 5251 Broad Branch Road, N.W., Washington DC 20015, USA
and Bayerisches Geoinstitut, University of Bayreuth,
Universitaetstrasse 30, D-95446 Bayreuth, Germany, e-mail:
razvan.caracas@uni-bayreuth.de}
\author{R.E. Cohen}
\affiliation{Geophysical Laboratory, Carnegie Institution of
Washington, 5251 Broad Branch Road, N.W., Washington DC 20015, USA}

\begin{abstract}
Using a {\it materials by design} approach, we find a class of
ordered oxynitride piezoelectrics with perovskite structure. We
predict that ordered YSiO$_2$N and YGeO$_2$N are characterized by
large nonlinear optic responses and by some of the largest
polarizations known to-date.
\end{abstract}

\date{\today}
\maketitle

Most piezoelectrics (PZs) or polar dielectrics are oxides, and those
commonly used are Pb-bearing compounds. PZs transform mechanical
energy into electrical energy and vice versa, and serve many uses
ranging from medical ultrasound and actuators to non-linear optic
devices \cite{Arnau04}. Materials for high temperature and/or high
electric field applications are highly desirable, like Pb-free PZ
materials as mandated by European law.

We have applied a synthetic solid state chemistry approach using ab
initio density functional calculations
\cite{Franceschetti99,Erwin04,Baettig05,Dudiy06,Bilc06,Kang06} to
search for useful materials. Instead of an automated search of
hundreds of compositions, we used a chemical approach similar to
what would be done to experimentally search for materials. We built
the oxynitride structures starting from the ideal cubic ABO$_3$
perovskite structure by replacing in each unit cell one oxygen atom
by one nitrogen (Fig. 1). In this way we aim to drive a stronger
polar distortion due to the strong covalent bonding expected between
the octahedrally coordinated B-cation and N, as well as the higher
nominal ionic charge of N$^{3-}$ compared with O$^{2-}$. The
resulting structures are ordered and polar with tetragonal symmetry,
{\em P}4{\em mm} space group and five atoms per unit cell. In order
to balance the -7 valence of O$_2$N we tried several combinations of
trivalent and tetravalent cations on the respectively A and B sites:
YSiO$_2$N, YGeO$_2$N, InTiO$_2$N, GaTiO$_2$N, BiTiO$_2$N, YTiO$_2$N,
BiZrO$_2$N, YCO$_2$N, YSnO$_2$N and YZrO$_2$N.

We used the local density approximation (LDA) of the density
functional theory \cite{Hohenberg64,Kohn65} with a planewave basis
and pseudopotentials in the ABINIT implementation \cite{Abinit02}.
We employed Troullier-Martins pseudopotentials \cite{Troullier91}
with a 40 Ha kinetic energy cut-off. We computed the dynamical and
dielectric properties and the coupled response of the energy to the
stresses and electric fields using density functional perturbation
theory \cite{Baroni01,Gonze05,Hamman05}.

We relaxed the {\em P}4{\em mm} atomic positions and cell parameters
at zero pressure for each of the above compositions and checked the
insulating/metallic character computing the electron band structure
and the corresponding density of states. We found that InTiO$_2$N
and GaTiO$_2$N are metallic and all the other compounds are
insulators. We checked the dynamical stability for each composition
computing the phonon band dispersion and from all the combinations
of cations listed above only YSiO$_2$N and YGeO$_2$N were
dynamically stable.

\begin{figure}
\includegraphics[width=0.8\linewidth]{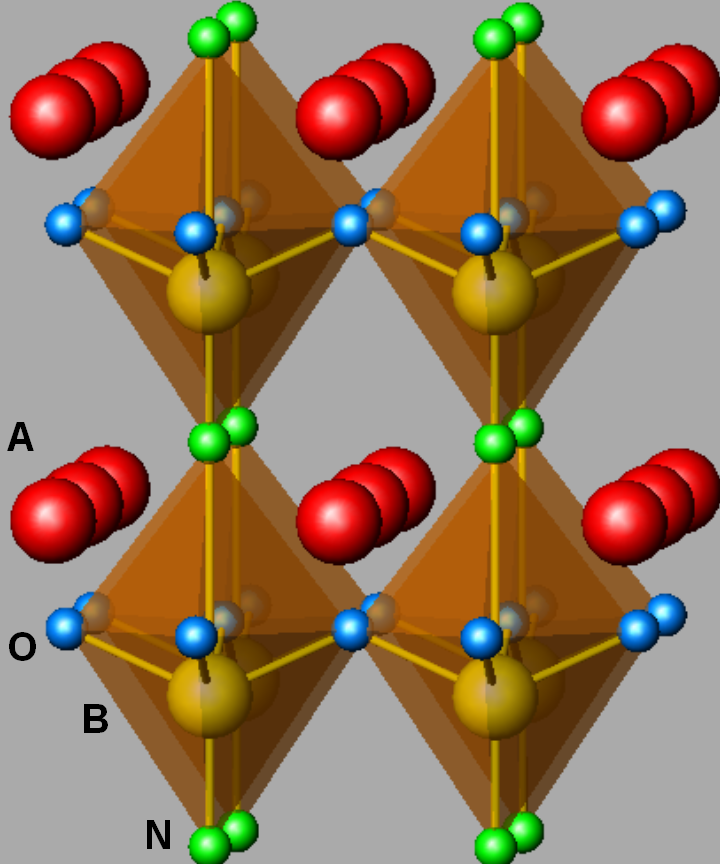}
\caption{Theoretical P4{\em mm} polar structure of the ordered
ABO$_2$N oxynitride perovskites. The most promising materials we
found have Y in the A site, and Si or Ge in the B site.}
\label{fig_struct}
\end{figure}

The crystal structure of the {\em P}4{\em mm} modification of
YSiO$_2$N and YGeO$_2$N is shown in Figure~\ref{fig_struct} and the
relevant structural parameters are listed in
Table~\ref{table_struct}. All the atoms are displaced relative to
their positions in the centrosymmetric structures. The corresponding
cation displacements are 0.76 \AA\ for Y and 0.61 \AA\ for Si in
YSiO$_2$N and 0.82 \AA\ for Y and 0.77 \AA\ for Ge in YGeO$_2$N.
These large displacements enhance the polar character of the
structures but also deepen the potential wells: 1.54 eV per unit
cell in YGeO$_2$N and 0.63 eV in YSiO$_2$N. Due to these large
potential wells, the polar structures will probably not be
switchable, thus these are not predicted to be ferroelectric without
modification.

Oxynitrides are generally produced by low-temperature
ammonization processes, which do not allow chemical ordering to
occur. \cite{Clarke02,Kim04,Marchand91,Hegde01}. We have tested
different atomic configurations and found that the ordered P4{\em
mm} structure is the lowest in energy, {\em e.g.} lower by at least
0.25 eV/O$_2$N group for YSiO$_2$N than any other disordered
configuration. We conclude that even if ordering in the oxynitrides
is difficult \cite{Ravel06}, the previous experimental structures
were disordered because of fabrication techniques. We suggest that
the ordered materials should be producible by high temperature
methods or by molecular beam epitaxy. Thus here we discuss ordered
polar materials.

\begin{figure}
\includegraphics[width=0.8\linewidth]{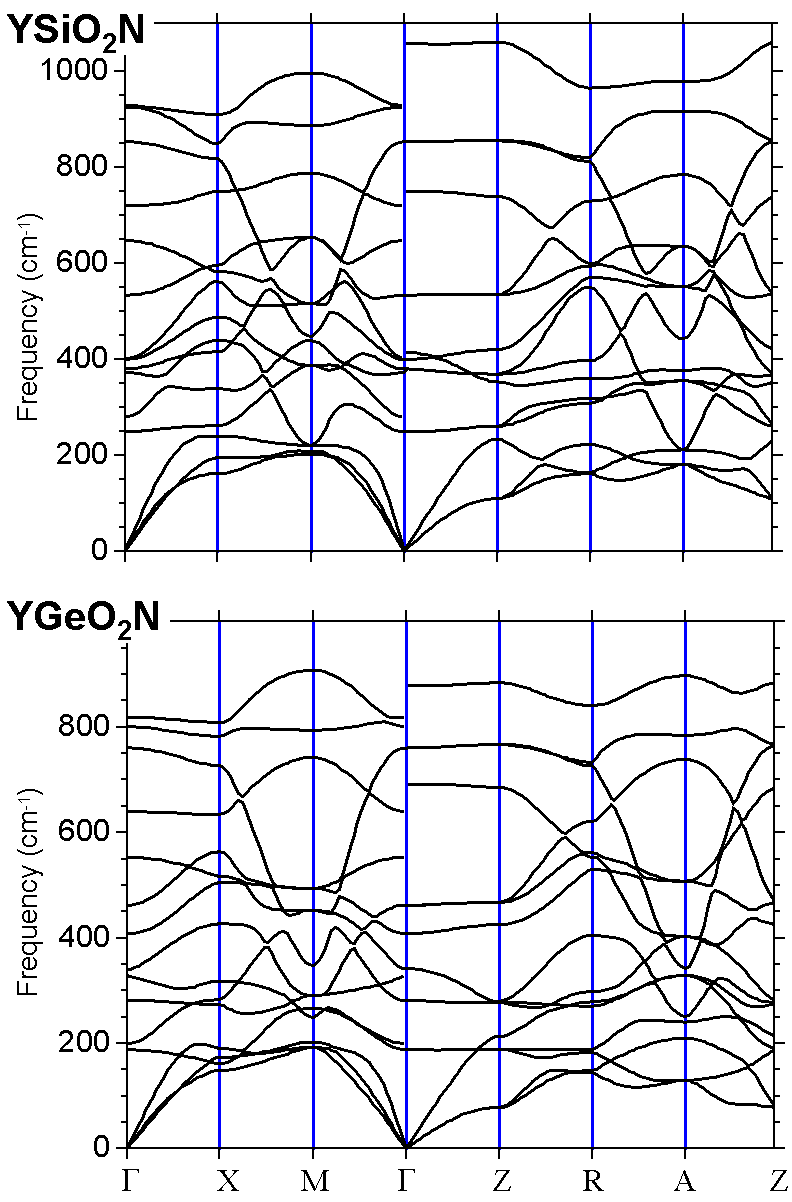}
\caption{Phonon band structure computed in the density functional
perturbation theory for YSiO$_2$N (a) and YGeO$_2$N (b). The path
through the Brillouin zone passes through: $\Gamma$=(0 0 0), X=(1/2
0 0), M=(1/2 1/2 0), Z=(0 0 1/2), R=(1/2 0 1/2), A=(1/2 1/2 1/2).}
\label{fig_phons}
\end{figure}

The phonon band dispersions for the two structures are shown in
Figure~\ref{fig_phons}. All the modes are stable, indicating
dynamical stability and thermodynamical (meta)stability. The elastic
constants also indicate mechanical stability
(Table~\ref{table_elastic}). The heat of formation of the
oxynitrides relative to a mixture of YN (fcc structure) and SiO$_2$
or GeO$_2$ (quartz structure) shows stability for the PZ phases
(-0.387 eV/unit cell for YSiO$_2$N and -0.309 for YGeO$_2$N). With
these stability criteria the PZ structures of the two oxynitrides
are at least thermodynamically metastable and their large negative
heat of formation suggests that they should be easily synthesized.

Both materials are insulators with large LDA band gaps, 2.6 eV in
YGeO$_2$N and 4.4 eV in YSiO$_2$N. The computed dielectric tensors
are $\epsilon^\infty$=[4.37 4.37 4.49] and $\epsilon^0$=[13.44 13.44
9.71] for YSiO$_2$N and $\epsilon^\infty$=[4.60 4.60 5.01] and
$\epsilon^0$=[16.10 16.10 9.80] for YGeO$_2$N. Because of symmetry,
the Born effective charges, $Z^{\alpha *}_{ij}=1/\Omega$ $\delta
P_i/\delta \tau^\alpha_j$, where $\Omega$ is the unit cell volume,
$P_i$ is the polarization along the direction $i$ and
$\tau^\alpha_j$ is the displacement of atom $\alpha$ along direction
$j$ at zero electric field, are diagonal.  Their values in the PZ
structures are listed in Table~\ref{table_born}. The Born effective
charges are only slightly enhanced due to the absence of the p-d
hybridization characteristic to other oxide perovskites
\cite{Cohen92,Posternak94}.

\begin{figure}
\includegraphics[width=1.0\linewidth]{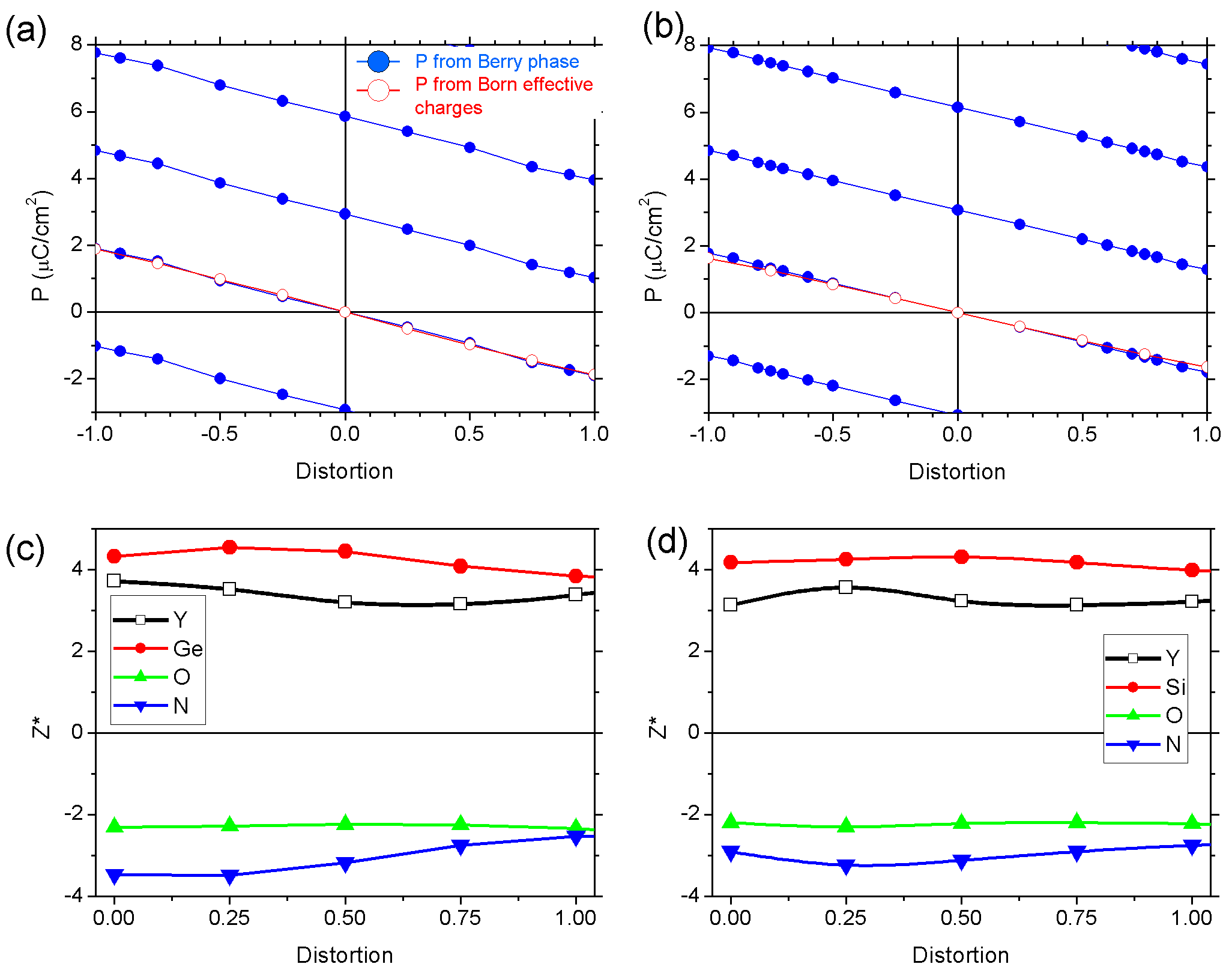}
\caption{The lattice of polarizations for YGeO$_2$N (a) and
YSiO$_2$N (b) and the Born effective charges for YGeO$_2$N (c) and
YSiO$_2$N (d) as a function of the position (=distortion) along the
centrosymmetric-to-ferroelectric direct path.} \label{fig_diel}
\end{figure}

We computed the spontaneous polarization using the modern theory of
polarization \cite{Kingsmith93,Vanderbilt94,Resta95}. The Berry's
phase polarization is a lattice of values, with a lattice spacing of
$2{\it e}{\bf R}/\Omega$, where ${\it e}$ is the electron charge and
{\bf R} are the lattice vectors. Polarization differences, however,
are unambiguous. Experiments measure polarization changes (usually
via hysteresis loops) and absolute polarizations are not measurable
using standard methods. For ferroelectrics with small polarizations
comparison between theory and experiment is straightforward, but in
materials with large polarizations the lattice of polarizations
measured for a single polar structure cannot be unambiguously
reduced to an experimentally meaningful value, called the {\it
effective} polarization, without additional computations
\cite{Neaton05,Resta07}. For tetragonal symmetry, only the
z-component of $P$ is non-zero. We obtained the formal polarization
along the polar axis {\em z} as P$_{Berry} = 130(\pm $n$308) \mu
C/cm^2$ for YSiO$_2$N and $103(\pm $n$293) \mu C/cm^2$ for
YGeO$_2$N, where the polarization lattice spacing is indicated in
parentheses. In order to find the effective polarization, we
computed its value using the Born effective charges, Z*
(Table~\ref{table_born}) and the displacement vectors from the ideal
perovskite structure, $\bf u$: $\Delta P=\sum\bf{Z^* u}$. This gives
effective polarizations P$_Z$=-163 $\mu$C/cm$^2$ for YSiO$_2$N and
P$_Z$=-171 $\mu$C/cm$^2$ for YGeO$_2$N. These values are not
identical to the Berry's phase because of the linear approximation,
but are very close. Using these values, we see that the effective
P$_{Berry}$ is -178 $\mu$C/cm$^2$ for YSiO$_2$N and -190
$\mu$C/cm$^2$ for YGeO$_2$N and these are our best estimates. The
direction of polarization is from the B ion towards the closer N.
Figure~\ref{fig_diel} shows the polarization lattices and comparison
of the effective polarization and the Berry's phase values. The Born
effective charges vary with distortion (Fig.~\ref{fig_diel}c and d),
the absolute values of the charges being slightly larger for the
centrosymmetric phase than the ferroelectric phase.

These values of the polarization are amongst the highest ever
reported so far in the literature. For comparison the spontaneous
polarization is  71 $\mu$C/cm$^2$ in LiNbO$_3$, 50 in
LiTaO$_3$,\cite{Lines01}, 70 in Pb(Mg,Nb)O$_3$ \cite{Kutnjak06} and
$>$150 for polycrystalline thin films of BiFeO$_3$\cite{Yun04}.
Other large polarization are those computed for pure PbTiO$_3$
(88)\cite{Szabo98}, BiFeO$_3$ (90-100)\cite{Neaton05}, PbVO$_3$
(150)\cite{Yoshitaka05}.

We computed the piezoelectric constants tensors,
$d_{ijk}=d\epsilon_{ij}/d\varepsilon_k$, where $\epsilon$ is the
stress tensor and $\varepsilon$ the electric field, using the linear
response theory of elasticity \cite{Hamman05,Wu05} and obtained
d$_{121}$=-12.5 pC/N, d$_{333}$=-8.3 pC/N and d$_{113}$=0.4 p/N
values for YSiO$_2$N and d$_{121}$=-20.5 pC/N, d$_{333}$=-5.5 pC/N
and d$_{113}$=-0.5 pC/N values for YGeO$_2$N. We also calculated the
non-linear optical coefficients, $d_{ijk}=\chi_{ijk}^{(2)}/2$ where
$\chi_{ijk}^{(2)}=3E^{\varepsilon_i\varepsilon_j\varepsilon_k}/V$ is
the third-order derivative of the energy with respect to electric
fields, via linear response \cite{Veithen05}, and obtained
d$_{113}$=2.03 pm/V and d$_{333}$=-5.5 pm/V in YSiO$_2$N and
d$_{113}$=2.63 pm/V and d$_{333}$=-4.51 pm/V in YGeO$_2$N. The
highest value to-date is shown by AANP, an organic material with
d$_{311}$=d$_{113}$=80 \cite{Tomaru91}. In KH$_2$PO$_4$, a commonly
used non-linear optical material, the non-linear optical
coefficients are on the order of 0.4 pm/V and in ADP on the order of
0.8 pm/V \cite{Eckardt90}. The electro-optic coefficients,
$c_{ijk}$, obtained from
$\Delta(\epsilon^{-1})_{ij}=\sum_k(c_{ijk}\varepsilon_k)$ where
$(\epsilon^{-1})_{ij}$ is the inverse of the dielectric tensor, have
values of c$_{133}$=-1.43 pm/V, c$_{333}$=0.64 pm/V and
c$_{331}$=-1.32 pm/V for YSiO$_2$N and c$_{133}$=-0.84 pm/V,
c$_{333}$=1.06 pm/V and c$_{331}$=-1.77 pm/V for YGeO$_2$N.

The high spontaneous polarizations, the non-linear optic
coefficients, and the large stability of the piezoelectric
structures make both YSiO$_2$N and YGeO$_2$N promising materials
with broad possible applications as piezoelectrics and non-linear
dielectrics. In addition, they may be useful as X-ray and neutron
generators via accelerations of ions at their surfaces
\cite{Naranjo05} due to their high polarizations.

\begin{acknowledgments}
Razvan Caracas acknowledges I. Grinberg for useful discussions. We
acknowledge support from the U.S. Office of Naval Research.
\end{acknowledgments}

\clearpage

\begin{table}
\caption{Structural parameters for YSiO$_2$N and YGeO$_2$N in the
P4{\em mm} ferroic structures. Y, Ge/Si, O and N atoms are in 1b(1/2
1/2 z), 1a(0 0 z), 2c(1/2 0 z) and 1a Wyckhoff positions
respectively.}\label{table_struct}
\begin{flushleft}
\begin{tabular}{lrrrrrrr}
\tableline
            & &  a(\AA)   &   c(\AA)   &   z$_{Y}$ &   z$_{Ge/Si}$ &   z$_{O}$ &   z$_{N}$ \\
YSiO$_2$N   & &  3.228   &   4.435   &   0.342   &   0.889   &   0.027   &   0.513   \\
YGeO$_2$N   & &  3.307   &   4.66    &   0.325   &   0.874   &   0.038   &   0.500   \\
\tableline
\end{tabular}
\end{flushleft}
\end{table}

\begin{table}
\caption{Elastic constants (in GPa) at zero pressure computed from
density functional perturbation theory
\cite{Baroni01,Hamman05}.}\label{table_elastic}
\begin{flushleft}
\begin{tabular}{lrrrrrrr}
\tableline

            &   C$_{11}$ &   C$_{33}$ &   C$_{12}$ &   C$_{13}$ &   C$_{44}$ &   C$_{66}$ \\
YSiO$_2$N   &   466 &   317 &   211 &   100 &   135 &   231 \\
YGeO$_2$N   &   374 &   345 &   176 &   86  &   90  &   167 \\

\tableline
\end{tabular}
\end{flushleft}
\end{table}

\begin{table}
\caption{Born effective charge tensors, Z$^*_{ij}$ in the
ferroelectric phases of YGeO$_2$N (columns 1-4) and YSiO$_2$N
(columns 5-8). Due to symmetry the tensors are diagonal. }
\label{table_born}
\begin{flushleft}
\begin{tabular}{l rrr l rrr}
\tableline
Elem. & Z$^*_{11}$ & Z$^*_{22}$ & Z$^*_{33}$ & Elem. & Z$^*_{11}$ & Z$^*_{22}$ & Z$^*_{33}$\\
\tableline

Y   &   3.48    &   3.48    & 3.38   &Y   &  3.47  &   3.47 &   3.22   \\
Ge  &   3.58    &   3.58    & 3.84   &Si  &  3.49  &   3.49 &   3.99    \\
O   &   -2.74   &  -1.91    & -2.34  &O   &  -2.66 &  -1.84 &   -2.22   \\
N   &   -2.40   &  -2.40    & -2.54  &N   &  -2.47 &  -2.47 &   -2.75  \\

\tableline
\end{tabular}
\end{flushleft}
\end{table}

\clearpage

\end{document}